\documentclass[a4,useAMS,usenatbib,usegraphicx]{mn2e}
\def\aapr{\ref@jnl{A\&A~Rev.}}		

\usepackage{amsmath}
\include{epsf}
\usepackage{color}
\usepackage{epsf}
\epsfclipon

\title[Cluster formation rates]{Formation rates of star clusters in the hierarchical merging scenario}
\author[R.Smith et al]{R.Smith$^{1}$\thanks{E-mail:rsmith@astro-udec.cl}, R. Slater${^1}$, M. Fellhauer${^1}$, S. Goodwin${^2}$, P. Assmann${^1}$\\
$^{1}$Departamento de Astronomia, Universidad de Concepcion, Casilla 160-C, Concepcion, Chile\\
\noindent
$^{2}$Department of Physics and Astronomy, University of Sheffield, Hicks Building, Hounsfield Road, Sheffield, S3 7RH, UK}
\begin{document}

\date{Accepted to MNRAS 10/05/11}

\pagerange{\pageref{firstpage}--\pageref{lastpage}} \pubyear{2011}

\maketitle

\label{firstpage}

\begin{abstract}
Stars form with a complex and highly structured distribution.  For a smooth
star cluster to form from these initial conditions, the star
cluster must erase this substructure. We study how substructure is
removed using N-body simulations that realistically handle two-body
relaxation. In contrast to previous studies, we find that hierarchical
cluster formation occurs chiefly as a result of scattering of stars
out of clumps, and not through clump merging. Two-body relaxation, in
particular within the body of a clump, can significantly increase the
rate at which substructure is erased beyond that of clump-merging
alone. Hence the relaxation time of individual clumps is a key
parameter controlling the rate at which smooth, spherical star
clusters can form. The initial virial ratio of the clumps is an
additional key parameter controlling the formation rate of a
cluster. Reducing the initial virial ratio causes a star cluster to
lose its substructure more rapidly.
\end{abstract}


\section{Introduction}

The vast majority of stars do not form alone. They appear to form in a
hierarchy in structures of tens to tens of thousands of stars 
(\citealp{testi00,gutermuth05,sanchez07,
andre07,goldsmith08,
andre10,
bressert10,difrancesco10,gutermuth09,
gouliermis10}).  Such structures are a natural consequence of
the gravo-turbulent model of star formation (e.g. \citealp{klessen00,
bonnell01, bonnell03,bate03, bonnell08, bate09, offner09}).

Hierarchical distributions are not in equilibrium, and will rapidly
dynamically evolve into dense star clusters or loose associations
(e.g. \citealp{aarseth72, Goodwin1998,bate98,boily99,kroupa03,goodwin04, 
allison09, allison10, moeckel09, Fellhauer2009,gieles10}).  
Such evolution is especially violent if the stars are initially dynamically cool 
(see \citealp{allison09,allison10}) as many observations suggest they are (e.g. \citealp{Walsh04,
difrancesco04,peretto06,walsh07,andre07,kirk07,gutermuth08b}).

In a clustered phase many
interesting processes may occur such as rapid dynamical mass
segregation (\citealp{allison09,allison10}), binary disruption and
modification (\citealp{heggie74,kroupa95,parker09}), the
formation of high-order multiples like the Trapezium system (\citealp{aarseth74,
zinnecker08,allison10}), and star-disc interactions
affecting planetary system formation (\citealp{boffin98,watkins98,pfalzner05,thies05,thies10}). 
Therefore an understanding of the collapse of hierarchical distributions is important
to understand the formation of star clusters and the possible
importance of these effects.

The evolution of initially substructured stellar distributions into
smooth star clusters has been studied by many authors (e.g. \citealp{aarseth72,
Goodwin1998,boily99,kroupa03,goodwin04,allison09,Fellhauer2009, 
moeckel09,Smith2011}).  In particular, \cite{Fellhauer2009}
attempted to quantify how rapidly an initially clumpy
distribution could evolve into a smooth star cluster.

In this paper we particularly extend the work of \cite{Fellhauer2009} 
and \cite{Smith2011} to investigate how a collapsing clumpy 
distribution in a static gas potential is able to erase its
substructure and form a smooth cluster.  \cite{Fellhauer2009}
presented a semi-analytic model for the erasure of substructure, but
they did not properly account for two-body effects in their
simulations.  Here we revisit their analysis with an accurate $N$-body
code. 

In Section~2 we present our initial conditions, in Section~3
we study the erasure of substructure before examining the rates at
which substructure is erased in Section~4.  Finally we discuss our
results in Section~5, and draw our conclusions in Section~6.

\begin{figure*}
\begin{center}$
\begin{array}{cc}
\includegraphics[height=2.4in,width=3.2in]{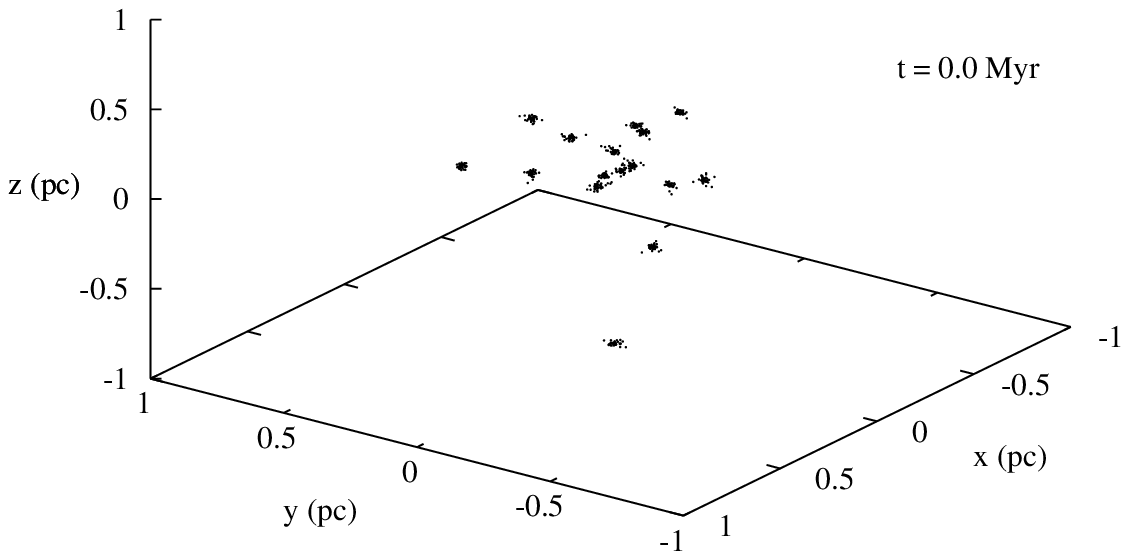} &
\includegraphics[height=2.4in,width=3.2in]{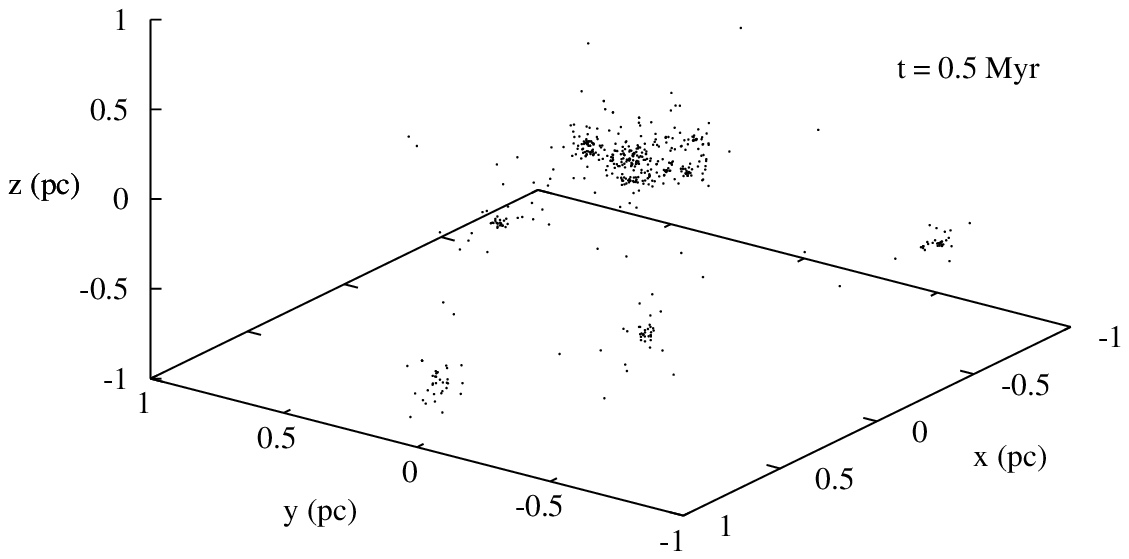}
\\ \includegraphics[height=2.4in,width=3.2in]{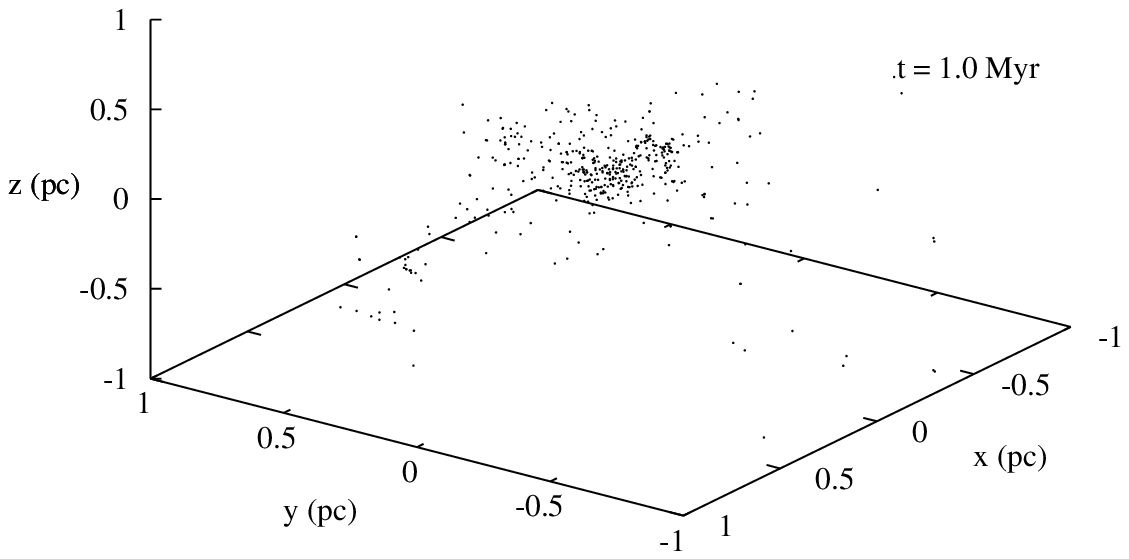} &
\includegraphics[height=2.4in,width=3.2in]{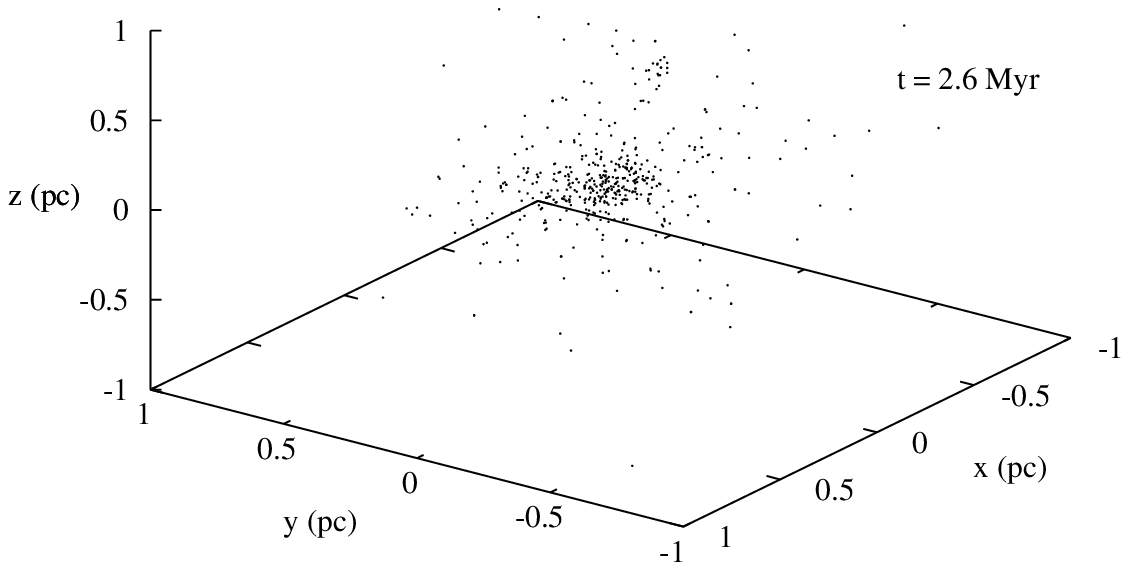} \\  
\end{array}$
\end{center}
\caption{An xyz-plot of evolution of substructure in a standard model
  simulation. The upper-left panel shows the initial stellar
  distribution (t=0.0 Myr). Stars are initially distributed in well
  defined clumps. The standard model star-forming region has a
  crossing-time $T_{\rm{cr}}^{\rm{sc}}=260$ kyr. As the simulation
  evolves, we show snapshots at 2 (upper-right panel), 4 (lower-left
  panel), and 10 (lower-right panel) crossing-times. Individual clumps
  `puff-up' significantly in less than 2 crossing-times. By 4
  crossing-times, only a few clearly defined clumps are visible, and by 10
  crossing-times almost all substructure has been erased.} 
\label{repsim}
\end{figure*}

\section{Initial conditions}

We perform our $N$-body simulations using the direct $N$-body integration
code {\sc{nbody6}} (\citealp{aarseth03}).  The advantage of {\sc{nbody6}} is that
it is able to rapidly and accurately model stellar dynamics, and
two-body encounters in particular.

Our initial conditions are similar to those of \citealp{Fellhauer2009}
(hereafter F09).  Our young star forming regions have a total mass
of $1000 M_\odot$.  We assume they convert gas to stars with an
efficiency $\epsilon$ which ranges between 0.1 and 0.8. Therefore the 
mass of stars is $\epsilon
\times 1000 M_\odot$, and the mass of gas $(1 - \epsilon) 
\times 1000 M_\odot$.  

We simulate the gas with a static background Plummer
potential with a Plummer scale radius of $R^{\rm{sc}}_{\rm{pl}}$ which
ranges between 0.02 and 1~pc.  We set a limiting cutoff radius for the
gas potential of $ 5 \times R^{\rm{sc}}_{\rm{pl}}$.  We
acknowledge that this is not ideal, as the gas will also dynamically
evolve.  However, it is presently impossible to simply model live
background gas and so we follow previous studies in including a static
background (e.g. \citealp{moeckel09}, F09 ,\citealp{Smith2011})

The stars are distributed within the gas potential in $N_0$ subclumps
which follow the underlying gas Plummer distribution.  $N_0$ ranges
from 4 to 32 resulting in a mass per clump of $M_{\rm pl} = 6$ to $80 M_\odot$ (where
the clump stellar mass is $M_{\rm pl} = (\epsilon \times 1000)/N_0
M_\odot$).

Subclumps are distributed within the Plummer sphere according to the 
prescription of \cite{aarseth74}.  Their bulk
velocities are then scaled to a desired virial ratio $Q_{\rm{i}} = T/|\Omega|$
(where $T$ is the total kinetic energy and $\Omega$ the total
potential energy) where $Q_{\rm{i}}=0.5$ is virial equilibrium and our scaling
ranges between $Q_{\rm{i}}=0$ and 0.5.

Each clump is assumed to be a virialised Plummer sphere with a Plummer scale
radius of $R_{\rm pl} = 0.01$~pc and a cut-off radius beyond which no
stars are placed of $R_{\rm cut} = 5 R_{\rm pl} =  0.05$~pc.  We assume
that sub-clumps are virialised initially as their relaxation time is
so short that they will rapidly virialise (but this process is also
effective at destroying clumps as we shall see).

We take equal-mass stars of mass $0.5 M_\odot$ (roughly the average
mass of a star from a standard IMF).  This means that our clumps
contain from 12 to 160 stars depending on the values of $\epsilon$ and
$N_0$.  Again we acknowledge that equal-mass stars are not realistic.
In particular, differences in stellar masses will have a significant
effect on two-body encounters which we are attempting to examine in
particular detail.  However, introducing a range of stellar masses
would significantly increase stochasticity and add another free
parameter, so we choose to ignore it for now.

Therefore the important parameters are:
\begin{itemize}
\item star formation efficiency $\epsilon$;
\item gas Plummer scale radius $R^{\rm{sc}}_{\rm{pl}}$;
\item number of subclumps $N_0$;
\item virial ratio of the stellar distribution $Q_{\rm{i}}$.
\end{itemize}
From these parameters it is possible to calculate a number of very
useful quantities.

The filling factor, $\alpha$, is the fraction of the volume which
contains subclumps
\begin{equation}
\alpha=\frac{R_{\rm{pl}}}{R^{\rm{sc}}_{\rm{pl}}}\rm.
\end{equation}

The crossing time of the whole system, $T^{\rm sc}_{\rm cr}$, or of an
individual clump, $T_{\rm cr}$ is the typical time taken to cross the
whole system or individual clump (the typical size divided by the
typical speed).

The two-body relaxation time is a measure of how rapidly the internal
velocities of a clump will change by order their own magnitude and is
given by

\begin{equation}
t_{\rm{relax}}=0.1 \frac{N_{\rm{part}}}{{\rm{ln}}(N_{\rm{part}})}
t_{\rm{cr}}\rm,
\end{equation}
where $N_{\rm{part}}$ is the total number of particles in the system,
and $t_{\rm{cr}}$ is the crossing-time of the system.

A list of all the simulations and their parameters can be seen 
in Table \ref{pars}.  Three different random realisations of each
parameter set are performed.

\begin{table*}
\centering
\begin{tabular}{|c|c|c|c|c|c|c|c|c|c|c|c|c|c|c|c|}
\hline $\alpha$ & $\epsilon$ & $N_0$ & $R^{\rm{sc}}_{\rm{pl}}$&
$R^{\rm{sc}}_{\rm{cut}}$ & $M^{\rm{sc}}_{\rm{pl}}$ &
$T^{\rm{sc}}_{\rm{cr}}$ & $Q_{\rm{i}}$ & $\sigma^{\rm{sc}}_{\rm{3D}}$
& $M_{\rm{star}}$ & $M_{\rm{gas}}$ & $R_{\rm{pl}}$ & $R_{\rm{cut}}$ &
$M_{\rm{pl}}$ & $T_{\rm{cr}}$ & t$_{\rm{relax}}$\\ & & & (pc) & (pc) &
(M$_\odot$) & (kyr) & & (km s$^{-1}$) & (M$_\odot$) & (M$_\odot$) &
(pc) & (pc) & (M$_\odot$) & (kyr) & (kyr)\\ \hline 0.05 & 0.32 & 16 &
0.20 & 1.00 & 1000 &  260 & 0.5 & 2.5 & 320 & 680 & 0.01 & 0.05 & 20.0
& 20 & 21.7 \\  0.05 & 0.32 & 16 & 0.20 & 1.00 & 1000 &  260 &
0.$\dot{3}$ & 2.0 & 320 & 680 & 0.01 & 0.05 & 20.0 & 20 & 21.7\\ 0.05
& 0.32 & 16 & 0.20 & 1.00 & 1000 &  260 & 0.1 & 1.4 & 320 & 680 & 0.01
& 0.05 & 20.0 & 20 & 21.7\\ 0.05 & 0.32 & 16 & 0.20 & 1.00 & 1000 &
260 & 0.0 & 0.0 & 320 & 680 & 0.01 & 0.05 & 20.0 & 20 & 21.7\\ \hline
0.01 & 0.32 & 16 & 1.00 & 5.00 & 1000 & 2950 & 0.5 & 1.1 & 320 & 680 &
0.01 & 0.05 & 20.0 & 20 & 21.7\\ 0.02 & 0.32 & 16 & 0.50 & 2.50 & 1000
& 1043 & 0.5 & 1.6 & 320 & 680 & 0.01 & 0.05 & 20.0 & 20 & 21.7\\ 0.10
& 0.32 & 16 & 0.10 & 0.50 & 1000 &   93 & 0.5 & 3.6 & 320 & 680 & 0.01
& 0.05 & 20.0 & 20 & 21.7\\  0.20 & 0.32 & 16 & 0.05 & 0.25 & 1000 &
33 & 0.5 & 5.0 & 320 & 680 & 0.01 & 0.05 & 20.0 & 20 & 21.7\\ 0.50 &
0.32 & 16 & 0.02 & 0.10 & 1000 &    8 & 0.5 & 8.0 & 320 & 680 & 0.01 &
0.05 & 20.0 & 20 & 21.7\\ \hline 0.05 & 0.32 &  4 & 0.20 & 1.00 & 1000
&  260 & 0.5 & 2.5 & 320 & 680 & 0.01 & 0.05 & 80.0 & 10 & 31.5
\\ 0.05 & 0.32 &  8 & 0.20 & 1.00 & 1000 &  260 & 0.5 & 2.5 & 320 &
680 & 0.01 & 0.05 & 40.0 & 15 & 27.4 \\ 0.05 & 0.32 & 32 & 0.20 & 1.00
& 1000 &  260 & 0.5 & 2.5 & 320 & 680 & 0.01 & 0.05 & 10.0 & 29 & 19.4
\\ \hline 0.05 & 0.10 & 16 & 0.20 & 1.00 & 1000 &  260 & 0.5 & 2.5 &
96 & 904 & 0.01 & 0.05 &  6.0 & 37 & 17.9 \\ 0.05 & 0.20 & 16 & 0.20 &
1.00 & 1000 &  260 & 0.5 & 2.5 & 200 & 800 & 0.01 & 0.05 & 12.5 & 26 &
20.2 \\ 0.05 & 0.25 & 16 & 0.20 & 1.00 & 1000 &  260 & 0.5 & 2.5 & 248
& 752 & 0.01 & 0.05 & 15.5 & 24 & 21.7 \\ 0.05 & 0.50 & 16 & 0.20 &
1.00 & 1000 &  260 & 0.5 & 2.5 & 496 & 504 & 0.01 & 0.05 & 31.0 & 17 &
25.5 \\ 0.05 & 0.60 & 16 & 0.20 & 1.00 & 1000 &  260 & 0.5 & 2.5 & 600
& 400 & 0.01 & 0.05 & 37.5 & 15 & 26.1 \\ \hline  0.10 & 0.10 & 16 &
0.10 & 0.50 & 1000 &   93 & 0.5 & 3.6 & 96 & 904 & 0.01 & 0.05 &  6.0
&  37 & 17.9 \\ 0.10 & 0.20 & 16 & 0.10 & 0.50 & 1000 &   93 & 0.5 &
3.6 & 200 & 800 & 0.01 & 0.05 & 12.5 &  26 & 20.2 \\ 0.10 & 0.25 & 16
& 0.10 & 0.50 & 1000 &   93 & 0.5 & 3.6 & 248 & 752 & 0.01 & 0.05 &
15.5 &  24 & 21.7 \\ 0.10 & 0.50 & 16 & 0.10 & 0.50 & 1000 &   93 &
0.5 & 3.6 & 496 & 504 & 0.01 & 0.05 & 31.0 &  17 & 25.5 \\ 0.10 & 0.70
& 16 & 0.10 & 0.50 & 1000 &   93 & 0.5 & 3.6 & 696 & 304 & 0.01 & 0.05
& 43.5 &  14 & 26.1 \\ \hline 0.05 & 0.80 & 32 & 0.20 & 1.0 & 1000 &
260 & 0.5 & 2.5 & 800 & 200 & 0.01 & 0.05 & 25.0 &  18 & 22.9 \\ 0.05
& 0.40 & 16 & 0.20 & 1.0 & 1000 &   260 & 0.5 & 2.5 & 400 & 600 & 0.01
& 0.05 & 25.0 &  18 & 22.9 \\ 0.05 & 0.20 & 8 & 0.20 & 1.0 & 1000 &
260 & 0.5 & 2.5 & 200 & 800 & 0.01 & 0.05 & 25.0 &  18 & 22.9
\\     \hline
\end{tabular}
\caption{A complete list of the parameters of all the simulations in
  our parameter study. The table is split by horizontal lines into
  sets (set 1 to set 6 from top to bottom). Each set is chosen to test
  the influence of a specific parameter on the formation rate of the
  cluster (see text for further details). Columns give the filling
  factor $\alpha$, the SFE $\epsilon$, number of clumps $N_0$,
  followed by the Plummer radius $R^{\rm{sc}}_{\rm{pl}}$, the cut-off
  radius $R^{\rm{sc}}_{\rm{cut}}$, the total mass
  $M^{\rm{sc}}_{\rm{pl}}$, and the crossing time of the star-forming
  region $T^{\rm{sc}}_{\rm{cr}}$. The following two columns are
  initial Virial ratio $Q_{\rm{i}}$, and corresponding velocity
  dispersion of clumps with respect to their clumps within the region
  $\sigma^{\rm{sc}}_{\rm{3D}}$. The next two columns are the mass in
  stars $M_{\rm{star}}$ and mass in gas $M_{\rm{gas}}$ (modelled as an
  analytical background) within the star-forming region. Finally we
  show the Plummer radius $R_{\rm{pl}}$, the cut-off radius
  $R_{\rm{cut}}$, the mass $M_{\rm{pl}}$, the crossing time
  $T_{\rm{cr}}$, and the relaxation time of an individual clump
  $t_{\rm{relax}}$.}
\label{pars}
\end{table*}

%%%%%%%%%%%%%%%%%%%%%%%%%%%%%%%%%%%%%%%%%%%%%%%%%%%%%%%%%%%%%%%%%%%%%%%%%%%%
\section{Results}
\label{results}

First we shall examine in detail our `standard model'.  This is a
virialised $Q_{\rm{i}} = 0.5$ cluster with
a star formation efficiency of $\epsilon = 0.32$ and $N_0 = 16$.  This
cluster has a stellar mass of $320 M_\odot$ in 16 $20 M_\odot$ clumps
(40 equal-mass stars per clump).  The cluster has a filling factor of
$\alpha = 0.05$ and a total crossing time of 260~kyr.  Each clump
has a crossing time of 20~kyr, and a relaxation time of 22~kyr.

Figure~\ref{repsim} shows the evolution of the standard model for 10
crossing times (2.6~Myr).  Initially (t=0 Myr) the stars have a highly clumpy and
sub-structured distribution. By 0.5~Myr the initial structure is
already less obvious, and by $>1$~Myr the cluster has a fairly smooth
appearance with very little evidence of the initial clumps.

\begin{figure*}
\begin{center}$
\begin{array}{cc}
\includegraphics[height=3.2in,width=3.2in]{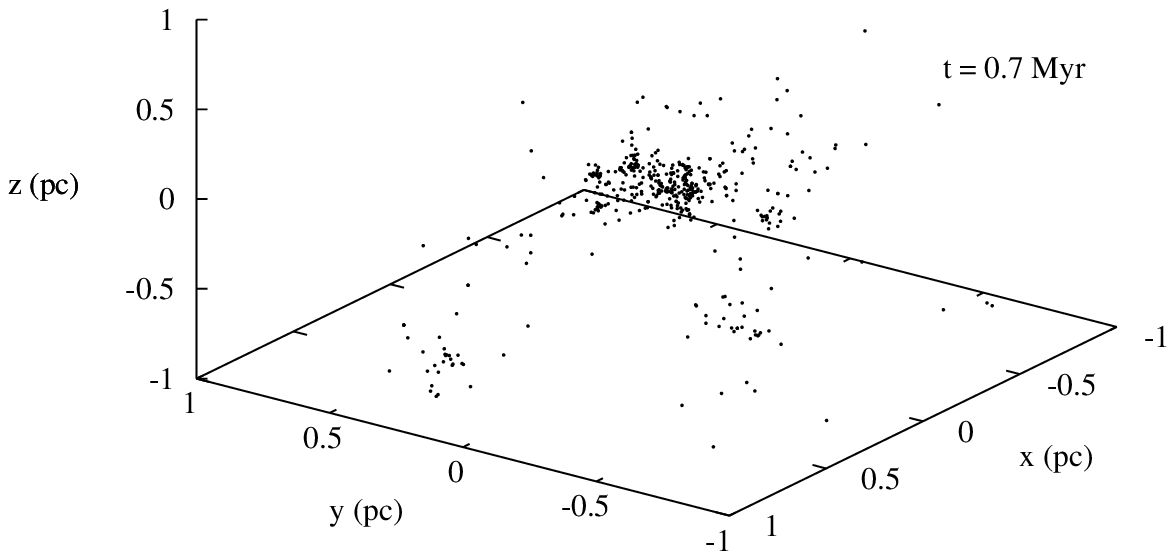} &
\includegraphics[height=3.2in,width=3.2in]{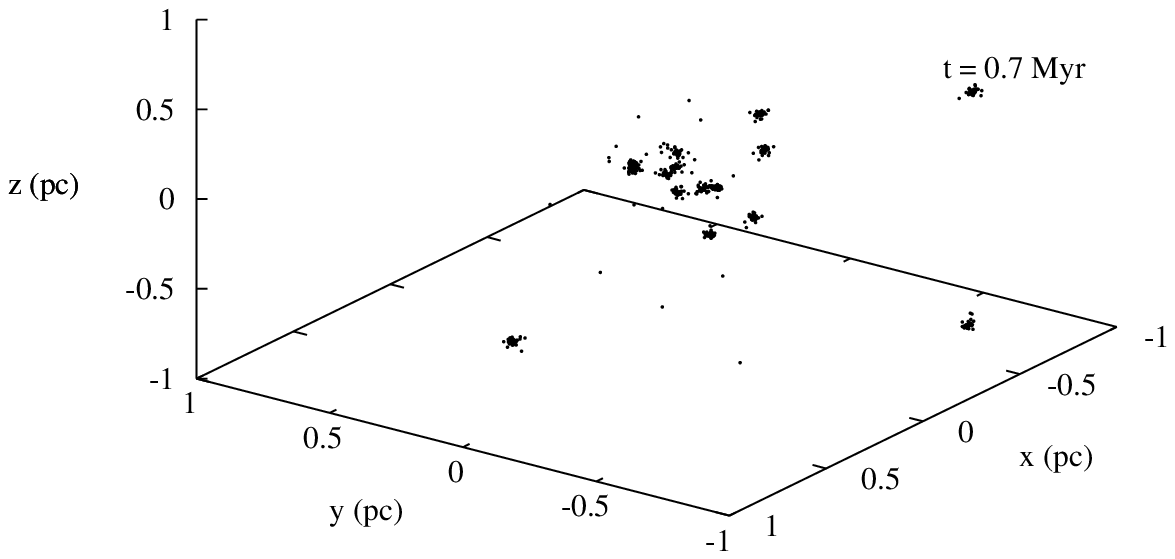} \\ 
\end{array}$
\end{center}
\caption{The resulting stellar distribution after 0.7 Myr of evolution
  from the standard model initial conditions in; an {\sc{nbody6}} simulation
  (left panel), and a {\sc{superbox}} simulation (right panel). Substructure
  has been erased more rapidly in the {\sc{nbody6}} simulations as a result
  of a realistic handling of two-body relaxation.}
\label{nbody6vssbox}
\end{figure*}

It is expected that the erasure of substructure would be due to one,
some, or all of the following mechanisms.
\begin{enumerate}
\item Internal scattering and the ejection of stars from a clump by internal
  two-body interactions.
\item Tidal stripping of clumps by the gas potential.
\item Tidal encounters between clumps. 
\item Collisions between clumps or stars.
\end{enumerate}

An examination of Figure~\ref{repsim} suggests that internal
scattering is a crucial factor in the erasure of substructure.  By
0.5~Myr, the appearance of the cluster is already quite smooth (we
shall return later to quantify the erasure of substructure, but for now we
will use a `by eye' examination).  In only two system crossing times
interactions between clumps cannot have been important.  The high
density of the initial clumps also suggests that tidal stripping by
other clumps or the gas potential can not have been responsible for the
smoothness.  The only process that works so rapidly on clumps is
internal relaxation as the clumps are 20 internal relaxation times old
by 0.5~Myr.

The effect of internal relaxation is to eject stars from the clumps
forming a smooth background of stars, and also to increase the size of 
the clumps (`puffing-up').  The puffing-up of clumps is obvious in
Figure~\ref{repsim} where even isolated clumps at 0.5~Myr are clearly
larger than initially. Binary formation within clumps is observed. This likely plays a significant role in enhancing scattering and ejection of stars from clumps.

As well as introducing a background of ejected stars, puffing-up has
two effects which further increase the rate of clump mergers.  As
clumps are larger they are more susceptible to tidal stripping, and
the filling factor increases -- significantly increasing the rate of
clump collisions.

In the clump merger simulations of F09 these effects arising from
two-body encounters were missed as the {\sc{superbox}} code used for these simulations
damps two-body encounters entirely.  This is illustrated in Figure
\ref{nbody6vssbox} showing two simulations with the same initial
conditions\footnote{It should be noted that the {\sc{superbox}} runs have
  $10^5$ much lower-mass particles initially in each clump and here we
  display only 40 for a fair comparison.}  having evolved for 0.7 
Myr (2.7 crossing-times). Therefore it is worth revisiting
the results of F09 when applied to star clusters in light of this 
vitally important physical process.  We note that the F09 results
apply to situations where two-body encounters can be considered
negligible.

\begin{figure}
  \centering \epsfxsize=9cm \epsfysize=7cm \epsffile{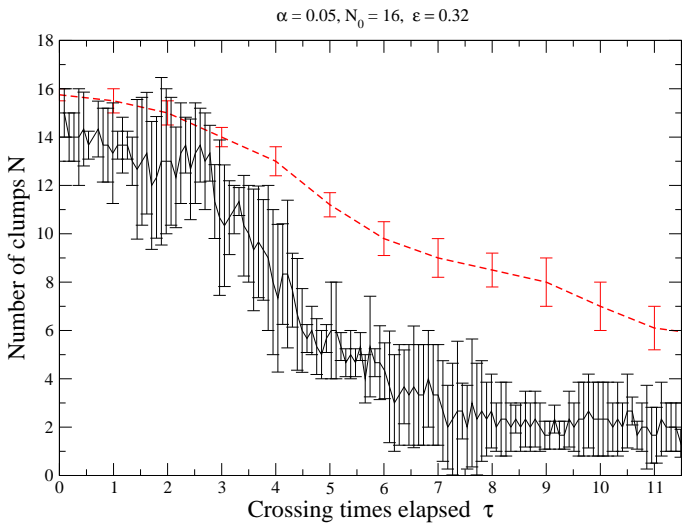}
  \caption{A graph of evolution of number of clumps N with time $\tau$
    (in units of the star-forming region crossing time). Both curves
    are results for simulations with the standard model initial
    conditions. The solid (black) line is the results for the {\sc{nbody6}}
    run as measured using a minimum spanning tree based clump
    finder. The dashed (red) line is the results for the {\sc{superbox}}
    run. As can be qualitatively assessed by eye in Figure
    \ref{nbody6vssbox}, substructure is more quickly erased in the
    {\sc{nbody6}} simulations as a result of a realistic handling of two-body
    relaxation. Equivalently, the {\sc{nbody6}} star cluster forms more
    rapidly.}
  \label{set1comp}
\end{figure}

\subsection{Clump counting with a Minimum Spanning Tree}

A minimum spanning tree (MST) is the shortest path linking a set of
$n$ points with no closed loops.  The MST is useful in that it always
has $n-1$ connections (edges) with a unique total length (see e.g. \cite{cartwright04}
 or \cite{allison09} for uses of the MST in 
astronomy).  We use the algorithm described in \cite{allison09} to construct an 
MST for a simulation at some point in time.

In order to identify clumps we introduce a cutting length
$l_{\rm{cut}}$.  If an edge has a length greater than $l_{\rm{cut}}$
then it is removed and we examine the subset of connections remaining
after cutting the longest edges.  If a subset contains more than
one-third of the stars initially within a subclump then we define it
as a clump.  (Note that \cite{gutermuth09} use a similar
method, as do many friends-of-friends clump finders.)  However, we 
introduce an extra element as our clumps are not just
close in physical space, but in phase space.  Therefore we build our
MST in 6D phase space before applying the cut.

A significant problem with clump finding is that it can be very
sensitive to the cutting length used.  In our initial conditions we
have the luxury of knowing what clumps are present and where they
are.  This allows us to fine tune our cutting length to get the right
answer initially (a bad choice of cutting length can result in
garbage).  We also check a number of simulations by eye to see that
the structures selected as clumps are indeed clumps, and that no 
structure have been missed.  This is not ideal, but the best we can do
at the moment with no way of properly selecting a cutting length
(\cite{gutermuth09} propose a way of selecting the cutting length,
but this only works well when structures are distinct).
\newline
\indent An illustration of the method is given in Figure \ref{set1comp} which
quantifies the evolution of the {\sc{nbody6}} and {\sc{superbox}} simulations
illustrated in Figure \ref{nbody6vssbox}.  The solid lines shows the rapid
decrease in the number of clumps in the {\sc{nbody6}} simulation compared to
those in the {\sc{superbox}} simulation in the dashed line.  This agrees with
the by eye assessment of the far more rapid erasure of substructure in
the former simulation (Figure \ref{repsim}).

\section{A parameter study of cluster formation rates}

With a quantitative measure of the evolution of the substructure and a
better understanding and modelling of the physical processes behind
this erasure, we can revisit the study of F09 to quantify the rate at
which structure is erased.

F09 suggested that the rate of erasure of substructure measured by the
number of clumps at time $\tau$, $N(\tau)$ compared to the initial
number of subclumps $N_0$ could be well-fitted by an equation of the
form 
\begin{equation}
\label{expeqn}
N(\tau)=(N_0-1) {\rm{exp}}(-\eta \tau)+1 \rm,
\end{equation}
where $\eta$ is a free parameter that depends on the initial
conditions of the simulation.  A large value of $\eta$ corresponds to
a rapid loss of substructure and the rapid appearance of a smooth
cluster.  Therefore we refer to $\eta$ as the `cluster formation rate'.

\begin{figure}
  \centering \epsfxsize=9cm \epsfysize=7cm \epsffile{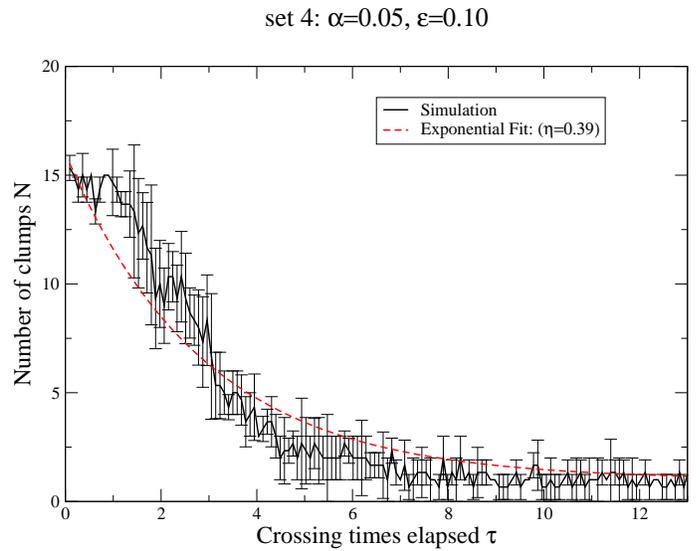}
  \caption{A graph of evolution of number of clumps N with time $\tau$
    (in units of the star-forming region crossing time). The solid
    (black) curve is the results of a Set 4 simulation as measured
    using the minimum spanning tree based clump finder. The dashed
    (red) line is an example of an exponential fit, using Equation
    \ref{expeqn}, to the simulation results. As presented in the key,
    this fit provides us with a value for the cluster formation rate
    parameter $\eta$. A high resulting value of $\eta$ indicates that
    substructure has been erased rapidly, and thus that a smooth
    cluster has formed rapidly.}
  \label{expfit}
\end{figure}

As shown in Figure \ref{expfit} we agree with F09 that an exponential
decay is indeed a good fit to the rate at which substructure is lost.

\subsection{Results of the parameter study on cluster formation rate}
\label{parstudyresults}

In Figure \ref{varypars} we show the variation of the cluster
formation rate $\eta$ with different parameters in our study (see
Table \ref{pars}).  Please note the differing y-axis scales in the
panels of Figure \ref{varypars}.  Remember also that each clump has the
same size in each simulation.

\subsection{The initial Virial ratio $Q_{\rm{i}}$}
The upper-left panel of Figure \ref{varypars} shows how cluster
formation rate depends on the initial virial ratio. As the initial
virial ratio becomes increasingly sub-virial (cool), the cluster formation rate
steadily increases. This occurs as a result of increased clump-clump
interactions - for sub-virial initial conditions clumps tend to fall
into a more compact configuration within the gas potential. Cluster
formation rate steeply rises as $Q_{\rm{i}}$ falls below 0.1. In this
case, clumps tend to fall towards the centre of the gas potential well
on time-scales $\sim1$ free-fall time, resulting in multiple
simultaneous clump-clump collisions. {\it{The initial Virial ratio
    $Q_{\rm{i}}$ is a key parameter controlling the cluster formation
    rate}}.

\subsection{The filling factor $\alpha$}
\begin{figure*}
\begin{center}$
\begin{array}{cc}
\includegraphics[height=6.8cm,width=9cm]{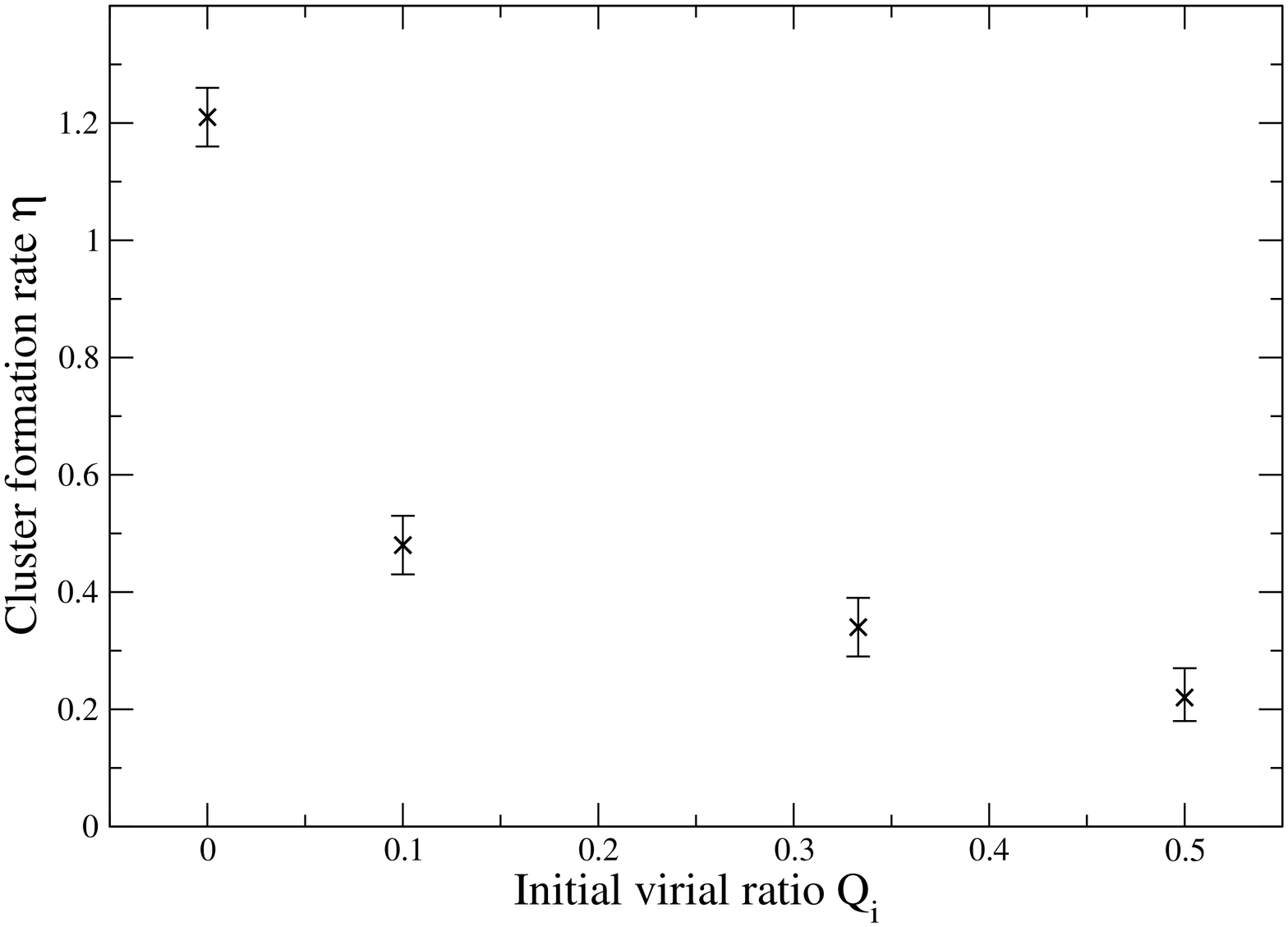} &
\includegraphics[height=6.8cm,width=9cm]{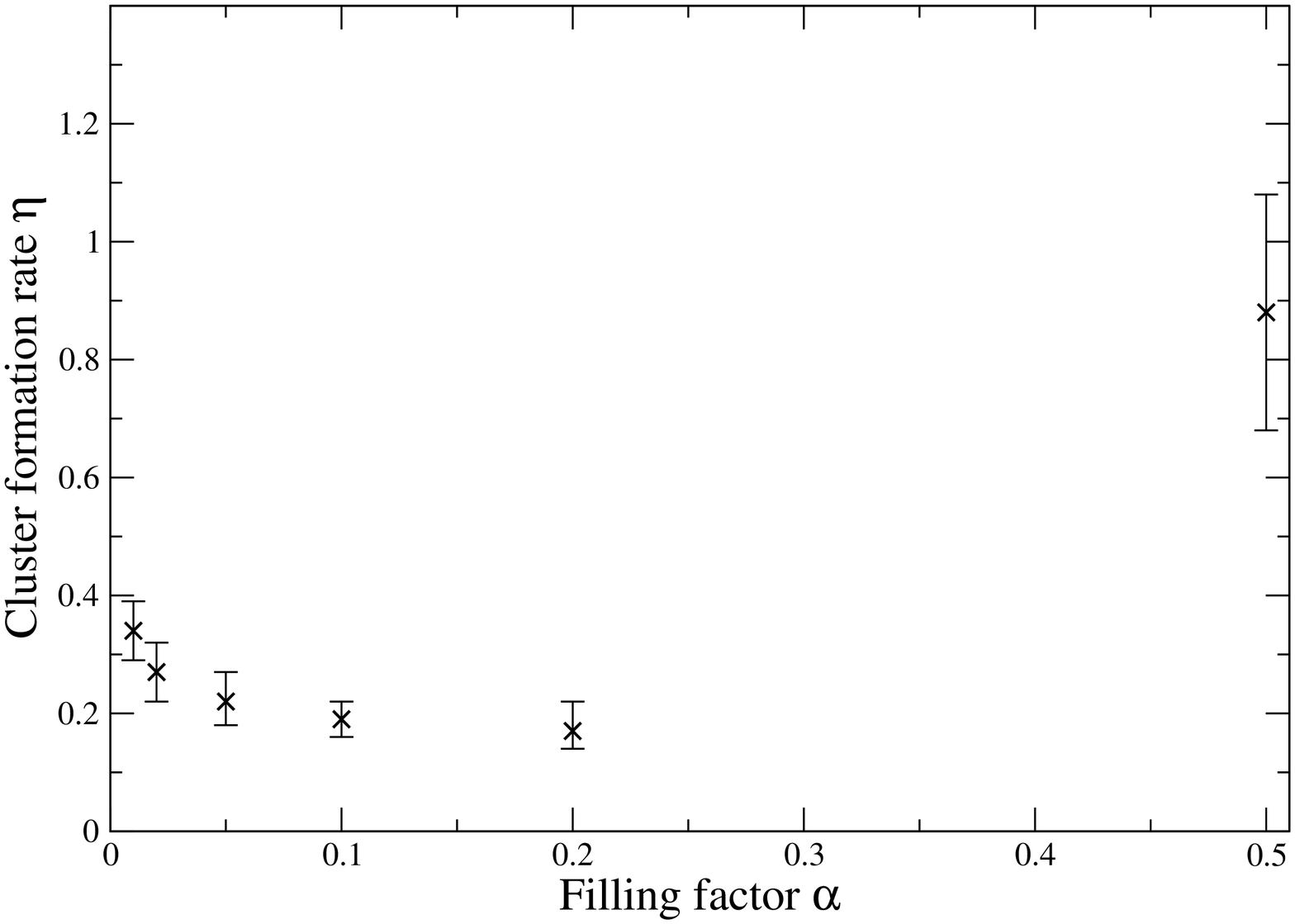}
\\ \\  \includegraphics[height=6.8cm,width=9cm]{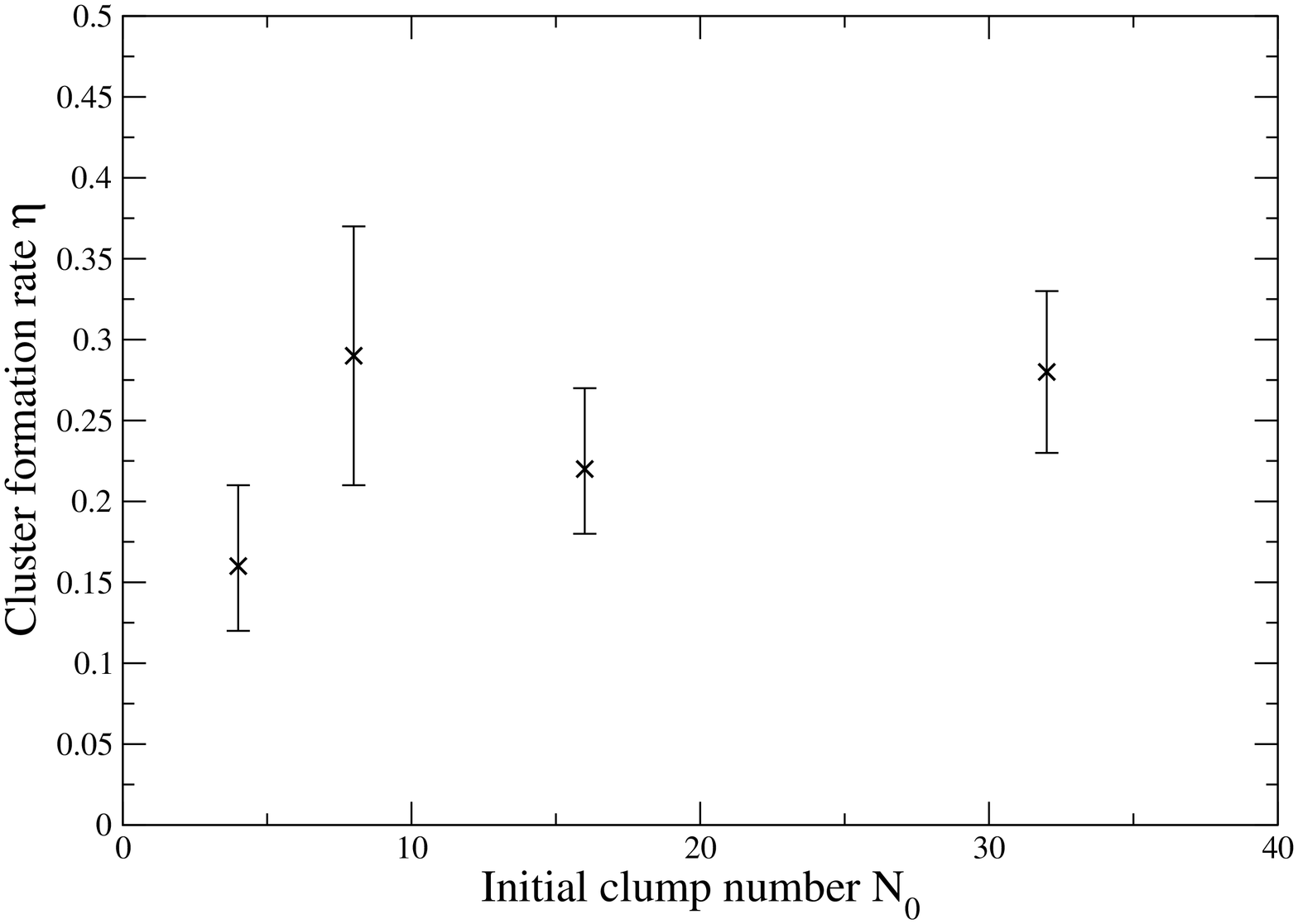} &
\includegraphics[height=6.8cm,width=9cm]{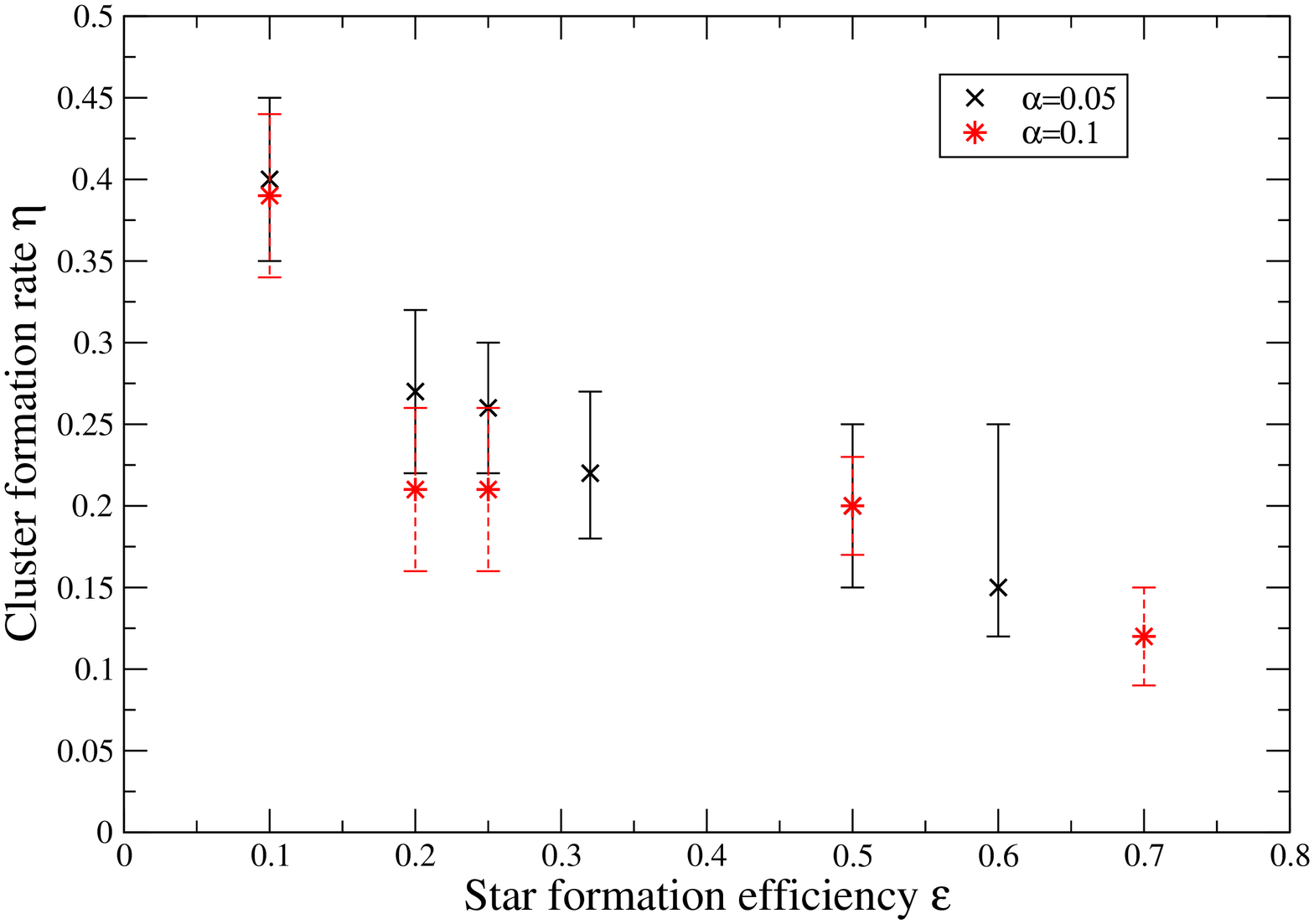}
\\ \\ \includegraphics[height=6.8cm,width=9cm]{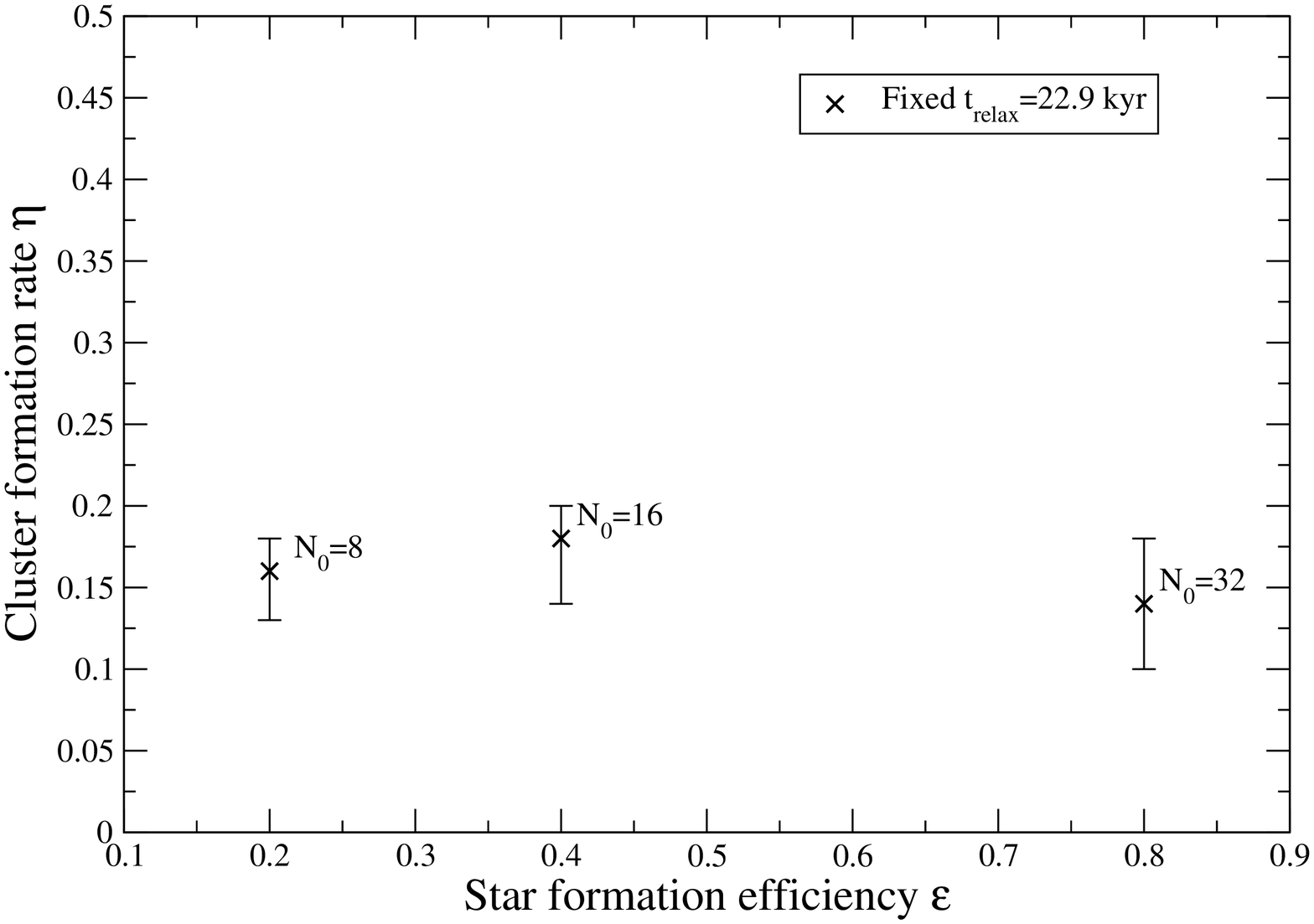} &
\includegraphics[height=6.8cm,width=9cm]{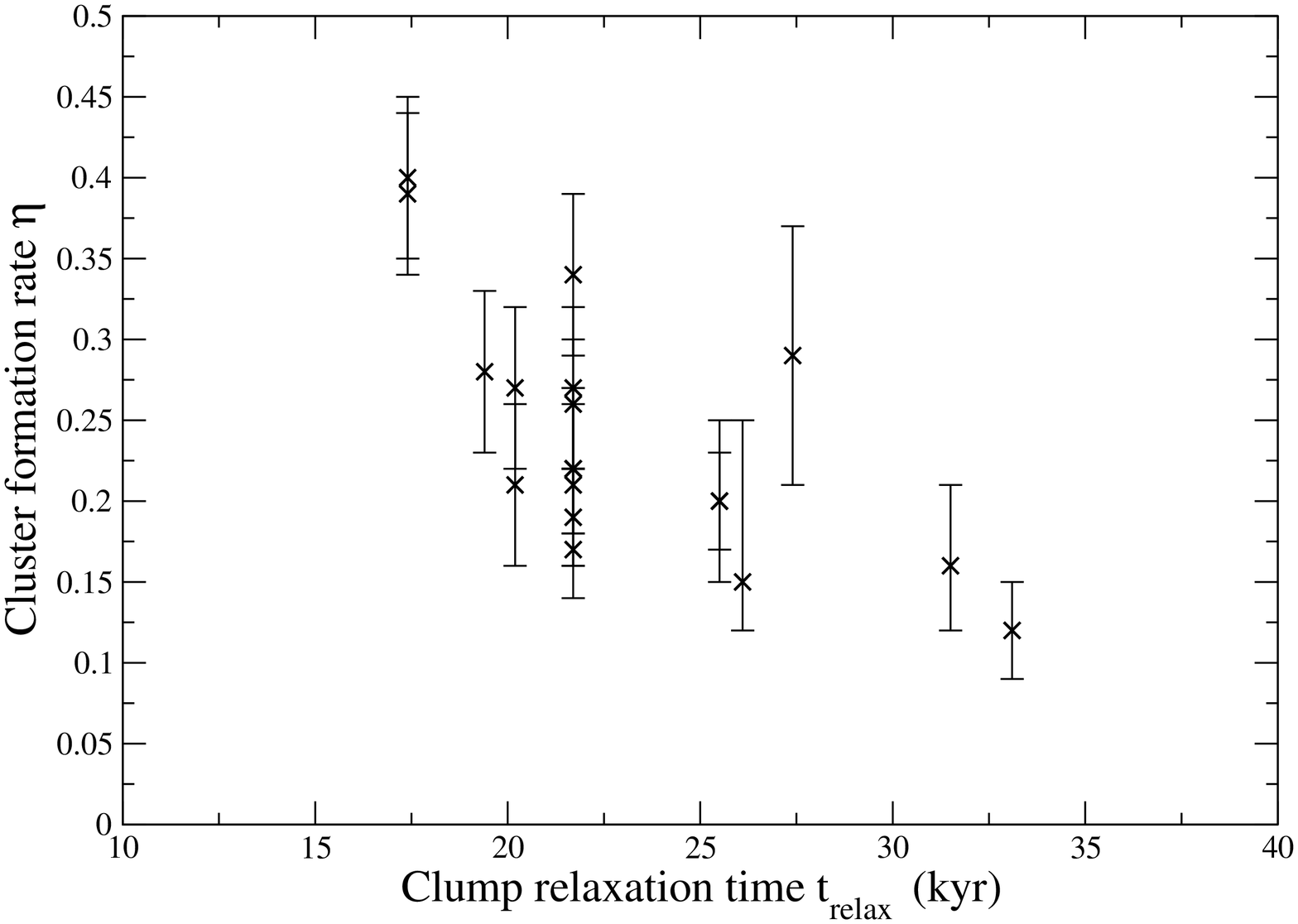} \\
\end{array}$
\end{center}
\caption{Plots of the dependency of the cluster formation rate
  parameter $\eta$ (y-axis) versus the parameters investigated in the
  parameter study (x-axis); initial Virial ratio $Q_{\rm{i}}$, filling
  factor $\alpha$ (upper-right), the initial number of clumps $N_0$
  (middle-left panel), region star formation efficiency $\epsilon$
  (middle-right panel), region star formation efficiency $\epsilon$
  for fixed clump relaxation time $t_{\rm{relax}}$ (bottom-left
  panel), and clump relaxation time $t_{\rm{relax}}$ (bottom-right
  panel). A description and discussion of these results is provided in
  the text (see Section \ref{parstudyresults}). Please note varying
  scale on y-axis. The two upper panels have a matching y-scale to
  each other. The four panels beneath them, all have matching y-scale
  to each other.}
\label{varypars}
\end{figure*}

The upper-right panel of Figure \ref{varypars} shows how cluster
formation rate depends on the filling factor $\alpha$. For
$\alpha<0.2$, the cluster formation rate is fairly constant as it is
dominated by the ejection of stars from clumps rather than clump-clump
interactions. However
for very high filling factors ($\alpha\sim0.5$) the cluster formation
rate becomes high. When filling factor is this high, clumps are
almost overlapping in the initial conditions, and this is enhanced by
clumps puffing-up, hence rapid merging
occurs. 

\subsection{The the initial number of clumps $N_0$ and star formation efficiency $\epsilon$}
The middle-left panel of Figure \ref{varypars} shows how the cluster
formation rate depends on the the initial number of clumps $N_0$, and the
middle-right panel shows how cluster formation rate depends on star
formation efficiency $\epsilon$.  At a first glance, the cluster formation
rate appears to depend on both $N_0$ and $\epsilon$. A trend towards
increasing cluster formation rate for increasing $N_0$ is visible in
the middle-left panel. Meanwhile, as star formation efficiency falls,
cluster formation rate increases. This trend is the reverse of
    what is seen in the F09 simulations. 

However by varying $N_0$ and $\epsilon$ we vary the initial number of
stars in a clump. This results in a variation in the relaxation time
$t_{\rm{relax}}$ of any individual clump and hence the speed at which
clumps puff-up (see Set 3 and Set 4 of Table
\ref{pars}). As $N_0$ increases, $t_{\rm{relax}}$ falls. Meanwhile as
$\epsilon$ increases, $t_{\rm{relax}}$ grows. Therefore it is
difficult to separate the additional effects of varying
$t_{\rm{relax}}$ on cluster formation rate. 

\subsection{A fixed clump relaxation time $t_{\rm{relax}}$}
In the bottom left panel of Figure \ref{varypars} we show how the
cluster formation rate varies for clumps with a fixed internal
relaxation time (selecting values of $N_0$ and $\epsilon$ such that
the relaxation time is 22.9~Kyr). For a constant $t_{\rm{relax}}$, the
cluster formation rate is roughly constant.  This strongly suggests
that it is the internal relaxation of clumps -- the ejection of stars
to make a background and the puffing-up of clumps to enhance
clump-clump collisions and tidal interactions that are the crucial
physical parameters.

\subsection{The relaxation time of individual clumps}
This conclusion is further supported by the 
bottom-right panel of Figure \ref{varypars} where the cluster
formation rate can be seen to decrease with greater clump internal
relaxation time.  Note that these simulations are all for an initially
virialised clump distribution, and we also exclude the $\alpha=0.5$
filling factor simulation as the cluster formation times in these
simulations are dominated by other effects.  It appears that {\it{the 
clump relaxation time is a key parameter controlling the cluster 
formation rate}}.

\section{Discussion}
\label{discussion}
As discussed in the introduction, star formation is a messy and complex
process that does not initially produce a smooth, relaxed star
cluster.  To give the roughly spherical, smooth star clusters that we
often observe, sub-structure must be erased (see also \citealp{aarseth72,
Goodwin1998,boily99,kroupa03,goodwin04,allison09,Fellhauer2009}). 

The speed at which a cluster can erase its substructure depends on its
initial virial ratio (see also \citealp{goodwin04}), but also
critically on the rate at which sub-structure evolves internally from
its initial state.  Dense, low-$N$ clumps have a short internal
relaxation time and will disperse rapidly (see also \citealp{kroupa03}).  
Two-body encounters within a clump eject stars, forming a
general stellar background, as well as causing clumps to increase
significantly in size, making clump-clump interactions more likely, and
making them more susceptible to tidal stripping. However, these
effects are secondary to internal relaxation.

By considering the middle-right
panel of Figure \ref{varypars} we can gauge the importance of clump
merging in our simulations. The data points are from Set 3 and Set 4
of our parameter study. By increasing the star formation efficiency,
we additionally increase the mass of the clumps. For clumps to merge,
their relative impact velocities must be of order the velocity
dispersion of the clumps. Therefore more massive clumps should merge
more easily as seen in the corresponding F09 simulations. Instead we
see the opposite - a decreasing cluster formation rate with increasing
clump mass. This indicates that merging plays a minor role in the
cluster formation process. Instead it is dominated by the effects of
inter-clump two-body encounters. For higher star formation
efficiencies, the initial number of stars within a clump
increases. Thus two-body encounters occur less frequently and
consequently the cluster formation rate falls (as seen in the
middle-right panel of Figure \ref{varypars}).

The theory of clump merging developed in \cite{Fellhauer2002} and F09
is therefore not applicable in low-$N$ ($N \sim 1000$) systems such as in
the star-forming regions modelled in this study. However, in high-$N$
systems such as mergers of clusters within cluster complexes, the 
effects of two-body encounters are far less important. In
these scenarios, the F09 theory should remain valid. 

An important conclusion to draw from this analysis is that clusters
can change their appearance rapidly when they are young.  A cluster
that appears smooth and relaxed at an age of a few~Myr may well not
have formed that way.  A number of authors have recently emphasised
that clusters evolve rapidly and that current conditions are not
always a good indicator of past conditions (e.g. \citealp{bastian08,
allison09}).

\section{Summary $\&$ Conclusions}
Both observations and theory agree that stars form with a complex
clumpy distribution within a star forming 
region. Small clumps containing
$\sim10$'s of stars form embedded within the envelope of molecular gas
from which they formed. However, observations of star clusters of $>$
a few~Myr in age often show smooth, relaxed distributions (e.g. the
Orion Nebula Cluster at $\sim 3$~Myr).  In order for a smooth 
spherical star cluster to
form, the star-forming region must erase its initial substructure.

We investigate the mechanisms by which substructure is erased by
modelling star forming regions using the {\sc{nbody6}} code. Our stars are initially
distributed in clumps, and embedded in a static potential to mimic the
gravitational influence of the gas envelope on stellar dynamics. We
conduct a parameter study to investigate the key parameters effecting
the rate at which substructure is erased. Key parameters include the
initial Virial ratio of the clumps within the gas potential, the
filling factor of clumps within the gas, the initial number of clumps,
and the star formation efficiency of the total star-forming region.

We find a number of new and different results, than of those presented
in a similar study in \citealp{Fellhauer2009} (F09). These differences
arise predominantly due to the proper treatment of two-body
encounters in our simulations. Our key results may be summarised in
the following:

\begin{enumerate}
\item Clusters form predominantly from stars that are scattered out of
  clumpy substructure, and not by clump merging.
\item As a result, the rate at which a cluster forms is a strong
  function of the relaxation time within the clumps. Unlike in F09,
  the star formation efficiency of the region does not effect the
  cluster formation rate.
\item The initial virial ratio of the clumps is also a key parameter
  controlling the rate at which a cluster forms. The lower the initial
  virial ratio, the more rapidly substructure is erased and a cluster
  forms.
\end{enumerate}

As inter-clump scattering has been demonstrated to be of such
importance to cluster formation, it is vital that models considering
the stellar dynamics of star-forming regions do so correctly. The use
of softened gravity between stars will result in suppression of
two-body encounters, and as such a key channel by which clusters form
will be missed.

Furthermore, if star clusters form by scattering of stars from clumps,
there is an increased likelihood that a moving subclump can leave a
trail of stars which maintain a velocity signature of the clump from
which they originated. Similar trails are reported in simulations of
massive stars clusters within a dark matter halo
(\citealp{Assmann2010}). Such velocity structures may be observable in
young clusters with the advent of Gaia. If so, we anticipate that
these observations could provide strong constraints on the recent
formation history of young clusters. We defer a detailed study of this
topic to a latter paper (\citealp{Smith2011} in prep.).

\section*{Acknowledgements}
MF announces financial support through FONDECYT grant 1095092. RS is financed by GEMINI-CONICYT fund 32080008 and a COMITE MIXTO grant. PA is financed through a CONCICYT PhD Scholarship.

\bibliography{bibfile}

\begin{thebibliography}{}

\bibitem[\protect\citeauthoryear{{Aarseth}}{{Aarseth}}{2003}]{aarseth03}
{Aarseth} S.~J.,  2003, {Gravitational N-Body Simulations}.
Cambridge University Press:~Cambridge, UK

\bibitem[\protect\citeauthoryear{{Aarseth}, {Henon} \& {Wielen}}{{Aarseth}
  et~al.}{1974}]{aarseth74}
{Aarseth} S.~J.,  {Henon} M.,    {Wielen} R.,  1974, \aap, 37, 183

\bibitem[\protect\citeauthoryear{{Aarseth} \& {Hills}}{{Aarseth} \&
  {Hills}}{1972}]{aarseth72}
{Aarseth} S.~J.,  {Hills} J.~G.,  1972, \aap, 21, 255

\bibitem[\protect\citeauthoryear{{Allison}, {Goodwin}, {Parker}, {Portegies
  Zwart} \& {de Grijs}}{{Allison} et~al.}{2010}]{allison10}
{Allison} R.~J.,  {Goodwin} S.~P.,  {Parker} R.~J.,  {Portegies Zwart} S.~F.,
   {de Grijs} R.,  2010, ArXiv e-prints

\bibitem[\protect\citeauthoryear{{Allison}, {Goodwin}, {Parker}, {Portegies
  Zwart}, {de Grijs} \& {Kouwenhoven}}{{Allison} et~al.}{2009}]{allison09}
{Allison} R.~J.,  {Goodwin} S.~P.,  {Parker} R.~J.,  {Portegies Zwart} S.~F.,
  {de Grijs} R.,    {Kouwenhoven} M.~B.~N.,  2009, \mnras, 395, 1449

\bibitem[\protect\citeauthoryear{{Andr{\'e}}, {Belloche}, {Motte} \&
  {Peretto}}{{Andr{\'e}} et~al.}{2007}]{andre07}
{Andr{\'e}} P.,  {Belloche} A.,  {Motte} F.,    {Peretto} N.,  2007, \aap, 472,
  519

\bibitem[\protect\citeauthoryear{{Andr{\'e}}, {Men'shchikov}, {Bontemps},
  {K{\"o}nyves}, {Motte}, {Schneider}, {Didelon}, {Minier} et~al.,}{{Andr{\'e}}
  et~al.}{2010}]{andre10}
{Andr{\'e}} P.,  {Men'shchikov} A.,  {Bontemps} S.,  {K{\"o}nyves} V.,  {Motte}
  F.,  {Schneider} N.,  {Didelon} P.,  {Minier} V.,    et~al., 2010, \aap, 518,
  L102

\bibitem[\protect\citeauthoryear{{Assmann}, {Fellhauer} \&
  {Wilkinson}}{{Assmann} et~al.}{2010}]{Assmann2010}
{Assmann} P.,  {Fellhauer} M.,    {Wilkinson} M.~I.,  2010, in {R.~de Grijs \&
  J.~R.~D.~L{\'e}pine} ed., IAU Symposium Vol.~266 of IAU Symposium, {Star
  clusters as building blocks for dSph galaxy formation}.
pp 353--356

\bibitem[\protect\citeauthoryear{{Bastian}, {Gieles}, {Goodwin}, {Trancho},
  {Smith}, {Konstantopoulos} \& {Efremov}}{{Bastian} et~al.}{2008}]{bastian08}
{Bastian} N.,  {Gieles} M.,  {Goodwin} S.~P.,  {Trancho} G.,  {Smith} L.~J.,
  {Konstantopoulos} I.,    {Efremov} Y.,  2008, ArXiv e-prints, 806

\bibitem[\protect\citeauthoryear{{Bate}}{{Bate}}{2009}]{bate09}
{Bate} M.~R.,  2009, \mnras, 397, 232

\bibitem[\protect\citeauthoryear{{Bate}, {Bonnell} \& {Bromm}}{{Bate}
  et~al.}{2003}]{bate03}
{Bate} M.~R.,  {Bonnell} I.~A.,    {Bromm} V.,  2003, \mnras, 339, 577

\bibitem[\protect\citeauthoryear{{Bate}, {Clarke} \& {McCaughrean}}{{Bate}
  et~al.}{1998}]{bate98}
{Bate} M.~R.,  {Clarke} C.~J.,    {McCaughrean} M.~J.,  1998, \mnras, 297, 1163

\bibitem[\protect\citeauthoryear{{Boffin}, {Watkins}, {Bhattal}, {Francis} \&
  {Whitworth}}{{Boffin} et~al.}{1998}]{boffin98}
{Boffin} H.~M.~J.,  {Watkins} S.~J.,  {Bhattal} A.~S.,  {Francis} N.,
  {Whitworth} A.~P.,  1998, \mnras, 300, 1189

\bibitem[\protect\citeauthoryear{{Boily}, {Clarke} \& {Murray}}{{Boily}
  et~al.}{1999}]{boily99}
{Boily} C.~M.,  {Clarke} C.~J.,    {Murray} S.~D.,  1999, \mnras, 302, 399

\bibitem[\protect\citeauthoryear{{Bonnell}, {Bate}, {Clarke} \&
  {Pringle}}{{Bonnell} et~al.}{2001}]{bonnell01}
{Bonnell} I.~A.,  {Bate} M.~R.,  {Clarke} C.~J.,    {Pringle} J.~E.,  2001,
  \mnras, 323, 785

\bibitem[\protect\citeauthoryear{{Bonnell}, {Bate} \& {Vine}}{{Bonnell}
  et~al.}{2003}]{bonnell03}
{Bonnell} I.~A.,  {Bate} M.~R.,    {Vine} S.~G.,  2003, \mnras, 343, 413

\bibitem[\protect\citeauthoryear{{Bonnell}, {Clark} \& {Bate}}{{Bonnell}
  et~al.}{2008}]{bonnell08}
{Bonnell} I.~A.,  {Clark} P.,    {Bate} M.~R.,  2008, \mnras, 389, 1556

\bibitem[\protect\citeauthoryear{{Bressert}, {Bastian}, {Gutermuth}, {Megeath},
  {Allen}, {Evans} II, {Rebull}, {Hatchell}, {Johnstone}, {Bourke}, {Cieza},
  {Harvey}, {Merin}, {Ray} \& {Tothill}}{{Bressert} et~al.}{2010}]{bressert10}
{Bressert} E.,  {Bastian} N.,  {Gutermuth} R.,  {Megeath} S.~T.,  {Allen} L.,
  {Evans} II N.~J.,  {Rebull} L.~M.,  {Hatchell} J.,  {Johnstone} D.,  {Bourke}
  T.~L.,  {Cieza} L.~A.,  {Harvey} P.~M.,  {Merin} B.,  {Ray} T.~P.,
  {Tothill} N.~F.~H.,  2010, arXiv:1009.1150

\bibitem[\protect\citeauthoryear{{Cartwright} \& {Whitworth}}{{Cartwright} \&
  {Whitworth}}{2004}]{cartwright04}
{Cartwright} A.,  {Whitworth} A.~P.,  2004, \mnras, 348, 589

\bibitem[\protect\citeauthoryear{{Di Francesco}, {Andr{\'e}} \& {Myers}}{{Di
  Francesco} et~al.}{2004}]{difrancesco04}
{Di Francesco} J.,  {Andr{\'e}} P.,    {Myers} P.~C.,  2004, \apj, 617, 425

\bibitem[\protect\citeauthoryear{{di Francesco}, {Sadavoy}, {Motte},
  {Schneider}, {Hennemann}, {Csengeri}, {Bontemps}, {Balog} et~al.,}{{di
  Francesco} et~al.}{2010}]{difrancesco10}
{di Francesco} J.,  {Sadavoy} S.,  {Motte} F.,  {Schneider} N.,  {Hennemann}
  M.,  {Csengeri} T.,  {Bontemps} S.,  {Balog} Z.,    et~al., 2010, \aap, 518,
  L91

\bibitem[\protect\citeauthoryear{Fellhauer, Baumgardt, Kroupa \&
  Spurzem}{Fellhauer et~al.}{2002}]{Fellhauer2002}
Fellhauer M.,  Baumgardt H.,  Kroupa P.,    Spurzem R.,  2002, Celestial
  Mechanics and Dynamical Astronomy, 82, 113

\bibitem[\protect\citeauthoryear{{Fellhauer}, {Wilkinson} \&
  {Kroupa}}{{Fellhauer} et~al.}{2009}]{Fellhauer2009}
{Fellhauer} M.,  {Wilkinson} M.~I.,    {Kroupa} P.,  2009, \mnras, 397, 954

\bibitem[\protect\citeauthoryear{{Gieles}, {Sana} \& {Portegies
  Zwart}}{{Gieles} et~al.}{2010}]{gieles10}
{Gieles} M.,  {Sana} H.,    {Portegies Zwart} S.~F.,  2010, \mnras, 402, 1750

\bibitem[\protect\citeauthoryear{{Goldsmith}, {Heyer}, {Narayanan}, {Snell},
  {Li} \& {Brunt}}{{Goldsmith} et~al.}{2008}]{goldsmith08}
{Goldsmith} P.~F.,  {Heyer} M.,  {Narayanan} G.,  {Snell} R.,  {Li} D.,
  {Brunt} C.,  2008, \apj, 680, 428

\bibitem[\protect\citeauthoryear{{Goodwin}}{{Goodwin}}{1998}]{Goodwin1998}
{Goodwin} S.~P.,  1998, \mnras, 294, 47

\bibitem[\protect\citeauthoryear{{Goodwin} \& {Whitworth}}{{Goodwin} \&
  {Whitworth}}{2004}]{goodwin04}
{Goodwin} S.~P.,  {Whitworth} A.~P.,  2004, \aap, 413, 929

\bibitem[\protect\citeauthoryear{{Gouliermis}, {Schmeja}, {Klessen}, {de Blok}
  \& {Walter}}{{Gouliermis} et~al.}{2010}]{gouliermis10}
{Gouliermis} D.~A.,  {Schmeja} S.,  {Klessen} R.~S.,  {de Blok} W.~J.~G.,
  {Walter} F.,  2010, \apj, 725, 1717

\bibitem[\protect\citeauthoryear{{Gutermuth}, {Megeath}, {Myers}, {Allen},
  {Pipher} \& {Fazio}}{{Gutermuth} et~al.}{2009}]{gutermuth09}
{Gutermuth} R.~A.,  {Megeath} S.~T.,  {Myers} P.~C.,  {Allen} L.~E.,  {Pipher}
  J.~L.,    {Fazio} G.~G.,  2009, \apjs, 184, 18

\bibitem[\protect\citeauthoryear{{Gutermuth}, {Megeath}, {Pipher}, {Williams},
  {Allen}, {Myers} \& {Raines}}{{Gutermuth} et~al.}{2005}]{gutermuth05}
{Gutermuth} R.~A.,  {Megeath} S.~T.,  {Pipher} J.~L.,  {Williams} J.~P.,
  {Allen} L.~E.,  {Myers} P.~C.,    {Raines} S.~N.,  2005, \apj, 632, 397

\bibitem[\protect\citeauthoryear{{Gutermuth}, {Myers}, {Megeath}, {Allen},
  {Pipher}, {Muzerolle}, {Porras}, {Winston} \& {Fazio}}{{Gutermuth}
  et~al.}{2008}]{gutermuth08b}
{Gutermuth} R.~A.,  {Myers} P.~C.,  {Megeath} S.~T.,  {Allen} L.~E.,  {Pipher}
  J.~L.,  {Muzerolle} J.,  {Porras} A.,  {Winston} E.,    {Fazio} G.,  2008,
  \apj, 674, 336

\bibitem[\protect\citeauthoryear{{Heggie}}{{Heggie}}{1974}]{heggie74}
{Heggie} D.~C.,  1974, in {A.~A.~Wyller} ed., Stability of the Solar System and
  of Small Stellar Systems Vol.~62 of IAU Symposium, {The role of binaries in
  cluster dynamics}.
pp 225--229

\bibitem[\protect\citeauthoryear{{Kirk}, {Johnstone} \& {Tafalla}}{{Kirk}
  et~al.}{2007}]{kirk07}
{Kirk} H.,  {Johnstone} D.,    {Tafalla} M.,  2007, \apj, 668, 1042

\bibitem[\protect\citeauthoryear{{Klessen} \& {Burkert}}{{Klessen} \&
  {Burkert}}{2000}]{klessen00}
{Klessen} R.~S.,  {Burkert} A.,  2000, \apjs, 128, 287

\bibitem[\protect\citeauthoryear{{Kroupa}}{{Kroupa}}{1995}]{kroupa95}
{Kroupa} P.,  1995, \mnras, 277, 1522

\bibitem[\protect\citeauthoryear{{Kroupa} \& {Bouvier}}{{Kroupa} \&
  {Bouvier}}{2003}]{kroupa03}
{Kroupa} P.,  {Bouvier} J.,  2003, \mnras, 346, 343

\bibitem[\protect\citeauthoryear{{Moeckel} \& {Bonnell}}{{Moeckel} \&
  {Bonnell}}{2009}]{moeckel09}
{Moeckel} N.,  {Bonnell} I.~A.,  2009, \mnras, 396, 1864

\bibitem[\protect\citeauthoryear{{Offner}, {Hansen} \& {Krumholz}}{{Offner}
  et~al.}{2009}]{offner09}
{Offner} S.~S.~R.,  {Hansen} C.~E.,    {Krumholz} M.~R.,  2009, \apjl, 704,
  L124

\bibitem[\protect\citeauthoryear{{Parker}, {Goodwin}, {Kroupa} \&
  {Kouwenhoven}}{{Parker} et~al.}{2009}]{parker09}
{Parker} R.~J.,  {Goodwin} S.~P.,  {Kroupa} P.,    {Kouwenhoven} M.~B.~N.,
  2009, \mnras, 397, 1577

\bibitem[\protect\citeauthoryear{{Peretto}, {Andr{\'e}} \&
  {Belloche}}{{Peretto} et~al.}{2006}]{peretto06}
{Peretto} N.,  {Andr{\'e}} P.,    {Belloche} A.,  2006, \aap, 445, 979

\bibitem[\protect\citeauthoryear{{Pfalzner}, {Vogel}, {Scharw{\"a}chter} \&
  {Olczak}}{{Pfalzner} et~al.}{2005}]{pfalzner05}
{Pfalzner} S.,  {Vogel} P.,  {Scharw{\"a}chter} J.,    {Olczak} C.,  2005,
  \aap, 437, 967

\bibitem[\protect\citeauthoryear{{S{\'a}nchez}, {Alfaro} \&
  {P{\'e}rez}}{{S{\'a}nchez} et~al.}{2007}]{sanchez07}
{S{\'a}nchez} N.,  {Alfaro} E.~J.,    {P{\'e}rez} E.,  2007, \apj, 656, 222

\bibitem[\protect\citeauthoryear{{Smith}, {Fellhauer}, {Goodwin} \&
  {Assmann}}{{Smith} et~al.}{2011}]{Smith2011}
{Smith} R.,  {Fellhauer} M.,  {Goodwin} S.,    {Assmann} P.,  2011, ArXiv
  e-prints

\bibitem[\protect\citeauthoryear{{Testi}, {Sargent}, {Olmi} \&
  {Onello}}{{Testi} et~al.}{2000}]{testi00}
{Testi} L.,  {Sargent} A.~I.,  {Olmi} L.,    {Onello} J.~S.,  2000, \apjl, 540,
  L53

\bibitem[\protect\citeauthoryear{{Thies}, {Kroupa}, {Goodwin}, {Stamatellos} \&
  {Whitworth}}{{Thies} et~al.}{2010}]{thies10}
{Thies} I.,  {Kroupa} P.,  {Goodwin} S.~P.,  {Stamatellos} D.,    {Whitworth}
  A.~P.,  2010, \apj, 717, 577

\bibitem[\protect\citeauthoryear{{Thies}, {Kroupa} \& {Theis}}{{Thies}
  et~al.}{2005}]{thies05}
{Thies} I.,  {Kroupa} P.,    {Theis} C.,  2005, \mnras, 364, 961

\bibitem[\protect\citeauthoryear{{Walsh}, {Myers} \& {Burton}}{{Walsh}
  et~al.}{2004}]{Walsh04}
{Walsh} A.~J.,  {Myers} P.~C.,    {Burton} M.~G.,  2004, \apj, 614, 194

\bibitem[\protect\citeauthoryear{{Walsh}, {Myers}, {Di Francesco}, {Mohanty},
  {Bourke}, {Gutermuth} \& {Wilner}}{{Walsh} et~al.}{2007}]{walsh07}
{Walsh} A.~J.,  {Myers} P.~C.,  {Di Francesco} J.,  {Mohanty} S.,  {Bourke}
  T.~L.,  {Gutermuth} R.,    {Wilner} D.,  2007, \apj, 655, 958

\bibitem[\protect\citeauthoryear{{Watkins}, {Bhattal}, {Boffin}, {Francis} \&
  {Whitworth}}{{Watkins} et~al.}{1998}]{watkins98}
{Watkins} S.~J.,  {Bhattal} A.~S.,  {Boffin} H.~M.~J.,  {Francis} N.,
  {Whitworth} A.~P.,  1998, \mnras, 300, 1214

\bibitem[\protect\citeauthoryear{{Zinnecker}}{{Zinnecker}}{2008}]{zinnecker08}
{Zinnecker} H.,  2008, in {E.~Vesperini, M.~Giersz, \& A.~Sills} ed., IAU
  Symposium Vol.~246 of IAU Symposium, {On the Origin of the Orion Trapezium
  System}.
pp 75--76

\end{thebibliography}

\bsp

\label{lastpage}

\end{document}